\documentclass[pra,nobalancelastpage,twocolumn,superscriptaddress,nofootinbib]{revtex4}
\pdfoutput=1
\usepackage{graphicx}
\usepackage{amsmath}
\usepackage{amssymb}
\usepackage[english]{babel}

\newcommand{\avg}[1]{\left< #1 \right>}

\newcommand{\up}{\uparrow}
\newcommand{\down}{\downarrow}

\newcommand{\ket}[1]{| #1 \rangle}
\newcommand{\bra}[1]{\langle #1 |}
\newcommand{\s}[1]{\substack{#1}}

\newcommand{\defs}{:=}
\begin{document}
\title{Detecting two-site spin-entanglement in many-body systems with local particle-number fluctuations}

\author{Leonardo Mazza}
\address{NEST, Scuola Normale Superiore \& Istituto Nanoscienze-CNR, I-56126 Pisa, Italy}

\author{Davide Rossini}
\address{NEST, Scuola Normale Superiore \& Istituto Nanoscienze-CNR, I-56126 Pisa, Italy}

\author{Rosario Fazio}
\address{NEST, Scuola Normale Superiore \& Istituto Nanoscienze-CNR, I-56126 Pisa, Italy}
\address{Centre for Quantum Technologies, National University of Singapore, 117543, Singapore}
\author{Manuel Endres}
\email{manuel.endres@mpq.mpg.de }
\address{Max-Planck-Institut f\"ur Quantenoptik, D-85748 Garching, Germany}

\begin{abstract}
We derive experimentally measurable lower bounds for the two-site entanglement of the spin-degrees of freedom of many-body systems with local particle-number fluctuations. Our method aims at enabling the spatially resolved detection of spin-entanglement in Hubbard systems using high-resolution imaging in optical lattices. A possible application is the observation of entanglement generation and spreading during spin impurity dynamics, for which we provide numerical simulations. More generally, the scheme can simplify the entanglement detection in ion chains, Rydberg atoms, or similar atomic systems.
\end{abstract}

\maketitle

\section{Introduction}
The role of entanglement for the quantitative understanding of quantum many-body systems has been the topic of a large number of theoretical studies~\cite{Amico:2008,Calabrese:2009,Eisert:2010}. In contrast, the experimental detection of entanglement in quantum many-body systems is less developed, which currently hinders the establishment of more direct links between experiments and theory. So far, entanglement witnesses have been extracted from macroscopic properties or diffractive probes, such as magnetic susceptibilities~\cite{Brukner:2006, Amico:2008, Lanting:2014}, spin- or atom-number squeezing parameters~\cite{Sorensen:2001,Esteve:2008,SchleierSmith:2010,Gross:2011}, or time-of-flight imaging~\cite{Cramer:2010, Cramer:2013}. Further, experiments using controlled collisions in optical lattices indicated the generation of entangled cluster states~\cite{Jaksch:1999,Mandel:2003}. However, these experiments did not access the spatial dependence of entanglement measures which is crucial for observing some of the elementary properties of entanglement in many-body systems, such as area laws~\cite{Eisert:2010} or the dynamical generation and spreading of entanglement~\cite{Schuch:2008,Amico:2008,Daley:2012}.\\
A candidate for establishing a direct experiment-theory connection are quantum spin systems~\cite{Amico:2008,Calabrese:2009,Eisert:2010,Osterloh:2002}. Such spin Hamiltonians can effectively describe the low-energy physics of certain materials~\cite{Auerbach:1994,  Lee:2006, Christensen:2007, Sachdev:2008}, for which a local detection of entanglement seems challenging. However, recent atomic physics realizations of quantum spin systems, such as neutral atoms in optical lattices~\cite{Kuklov:2003, Duan:2003, Trotzky:2008a, Simon:2011,Fukuhara:2013, Fukuhara:2013b,Greif:2013,Nascimbene:2013} and trapped ions~\cite{porras:2004,Friedenauer:2008,Kim:2009, blatt:2012, schneider:2012, Jurcevic:2014, Richerme:2014}, offer the possibility of a local read-out of spin correlations. \\
In ion traps, such local detection of spin correlations and entanglement has been the standard for many years but was mostly used in the context of quantum computing~\cite{Haffner:2008}. Only recently, these techniques were employed to detect entanglement in a simulation of a spin system showing the first spatially-resolved detection of entanglement spreading after a local quantum quench~\cite{Jurcevic:2014}.\\
For quantum many-body systems in optical lattices, local detection of individual particles and their correlations has only been demonstrated in the past few years using high-resolution microscopy~\cite{Bakr:2009, Bakr:2010, Sherson:2010, Endres:2011,Simon:2011,Fukuhara:2013, Fukuhara:2013b}. Proposals have been made to detect the R\'{e}nyi entropy with this technique~\cite{Alves:2004, Daley:2012,Pichler:2013,Abanin:2012} but no experiment has shown the spatially resolved detection of entanglement in such systems to date.\\
A key difference between quantum magnetism experiments in ion traps and optical lattices is that in the latter, on-site number fluctuations coexist with spin fluctuations. The reason is that spin interactions in optical lattices are typically generated via superexchange as a second order process in the large interaction limit of Hubbard models~\cite{Kuklov:2003, Duan:2003,Trotzky:2008a}, where number fluctuations are suppressed but not absent.\\
Particularly in low dimensions, local number fluctuations can be sizable even at zero temperature~\cite{Bakr:2010,Endres:2011}, and additionally the currently achievable temperatures lead to thermal activation of defects~\cite{Sherson:2010}. In solids, number fluctuations naturally arise through hole-doping of Mott insulators, leading to an effective description in terms of $t$-$J$ models~\cite{Lee:2006}.\\
For such systems, a detection of spin-entanglement must take the presence of occupation number fluctuations into account. On the theoretical level, the distinction between entanglement in internal and number degrees of freedom has been clarified in Refs.~\cite{Wiseman:2003, Dowling:2006, Schuch:2004}. However, concrete experimental proposals for detecting the entanglement between spins in quantum many-body systems of atoms with local particle-number fluctuations are lacking.\\
Here we propose an experimentally feasible scheme to detect spin-entanglement between two sites in the presence of number fluctuations in Hubbard systems using single-atom- and single-site-resolved imaging of atoms in optical lattices~\cite{Bakr:2009, Bakr:2010, Sherson:2010}. To this end, the key challenges are current limitations in these setups, namely the lack of arbitrary local spin rotations~\cite{Weitenberg:2011}, the lack of full spin resolution~\cite{Fukuhara:2013}, and the parity-projection problem~\cite{Bakr:2010, Sherson:2010}. Fully accounting for these restrictions, we derive detectable lower bounds for the concurrence~\cite{Wooters:1998}, an entanglement measure, of the spin-degree of subsystems consisting of two lattice sites. Our method can be readily implemented in current high-resolution imaging setups for optical lattices without technical modifications~\cite{Bakr:2009, Bakr:2010,Simon:2011, Sherson:2010, Endres:2011,Fukuhara:2013}.\\
The scheme is immediately applicable to studying the entanglement generation and spreading during single spin-impurity dynamics in one-dimensional Bose-Hubbard chains~\cite{Fukuhara:2013,Subrahmanyam:2004, Amico:2004}. For this scenario, we provide numerical simulations identifying a parameter range where such experiments could be performed.\\
While our focus is on spin-impurity dynamics, the method can be used in a broader context. For example, it could be an important diagnostic tool in the current experimental search for antiferromagnetic order in the fermionic Hubbard model realized with cold gases~\cite{Esslinger:2010}. For ion trap implementations of quantum magnetism, the bounds derived in Sec.~\ref{spinresult} could lead to a simplified detection of entanglement in impurity dynamics~\cite{Jurcevic:2014} or global quantum quenches~\cite{Richerme:2014} without the need for a full state reconstruction. Further, our results also apply to experiments with Rydberg atoms in optical tweezers~\cite{Gaetan:2009,Urban:2009,Wilk:2010,Isenhower:2010,Gaetan:2010}, where atom number fluctuations can result from trap loss. Finally, our method could be used to detect the entanglement in spatially ordered structures of Rydberg excitations in optical lattices~\cite{Schauss:2012}.\\
The outline of the paper is as follows. In Sec.~\ref{single_impurity}, we give an introduction to entanglement generation and spreading during single impurity dynamics in the nearest neighbor spin-$1/2$ XX-chain~\cite{Subrahmanyam:2004, Amico:2004}. The derivation of lower bounds for the concurrence then follows in several steps taking into account the known experimental limitations for high-resolution imaging of quantum gases in optical lattices. In Sec.~\ref{spinresult}, we derive a lower bound neglecting number fluctuations based only on global pulses in order to circumvent the lack of arbitrary local spin rotations~\cite{Weitenberg:2011}. We then give a conceptual introduction to the detection of spin-entanglement in the presence of number fluctuations in Sec.~\ref{number_flctuations}, followed by a case study of spin impurity dynamics in the one-dimensional Bose-Hubbard model in Sec.~\ref{case}. We extend the detection scheme to include number fluctuations in Sec.~\ref{scheme_full_spin} assuming fully spin-resolved detection. In Sec.~\ref{nospin_resolution}, we account for the current inability to detect two different spin states at once~\cite{Fukuhara:2013} and also treat the restriction to local parity imaging~\cite{Bakr:2010,Sherson:2010}. We finish with a conclusion and outlook section.
\section{Entanglement during impurity dynamics in the XX-chain}\label{single_impurity}
\begin{figure}[t]
  \includegraphics[width=\columnwidth]{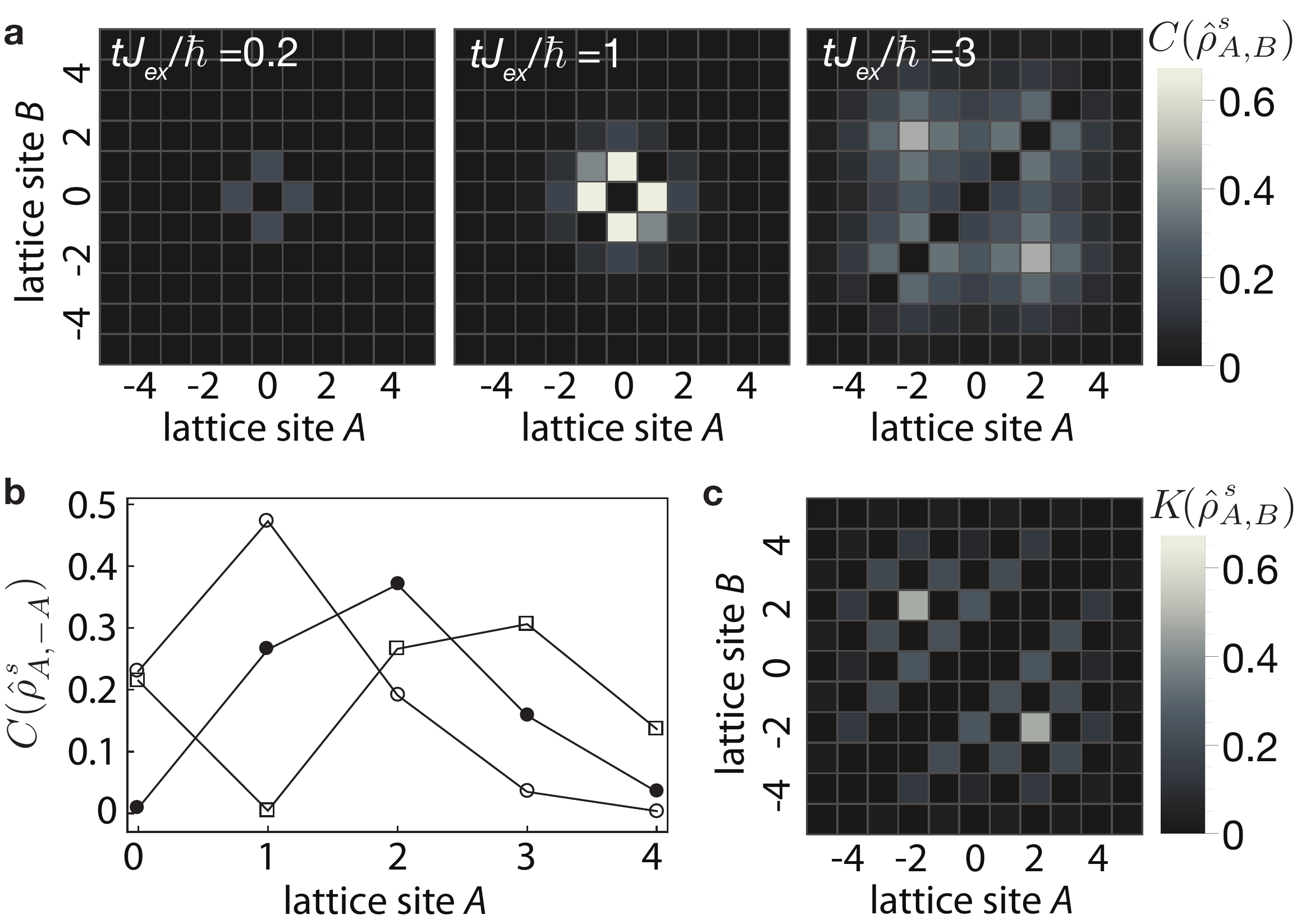}
  \caption{\textbf{a} Density plot of the concurrence $C(\hat \rho_{A,B}^s(t))$ for the single spin impurity dynamics as a function of lattice sites $A$ and $B$ for $tJ_{\rm ex}/\hbar=0.2, 1, 3$ (left, middle, right). \textbf{b} Concurrence $C(\hat \rho_{A,-A}^s(t))$ for the single spin impurity dynamics evaluated at sites $A,-A$ for $tJ_{\rm ex}/\hbar=3, 4, 5$ (open circles, filled circles, open rectangles). Lines are shown as a guide for the eye. \textbf{c} Density plot of the lower bound for the concurrence $K(\hat \rho_{A,B}^s(t))$ (see Eq.~(\ref{bound2})) for the single spin impurity dynamics as a function of lattice sites $A$ and $B$ for $tJ_{\rm ex}/\hbar=3$. Note that $K(\hat \rho_{A,B}^s(t))=0$ for odd distances $A-B$, which results in a checkerboard pattern.}
  \label{fig:Figure1}
\end{figure}

To provide a concrete example and target application, we review the entanglement generation and spreading during spin impurity dynamics in a \mbox{spin-1/2} XX-chain~\cite{Subrahmanyam:2004, Amico:2004} with Hamiltonian
\begin{align}
\hat{H}_{\rm XX}=-\frac{J_{\rm ex}}{2}\sum_j(\hat{S}_j^+\hat{S}_{j+1}^-+\hat{S}_{j+1}^+\hat{S}_{j}^-), \label{heisenberg}
\end{align}
where $J_{\rm ex}$ is the exchange coupling and $\hat{S}_j^\pm=\frac{1}{2}(\hat \sigma^x_j\pm i\hat\sigma^y_j)$ are spin-1/2 raising (lowering) operators. With $\hat \sigma^\alpha_j$ ($\alpha=x,y,z$) we denote the Pauli operators applied to site $j$.\\
Hamiltonians of this type are important for describing recent experiments realizing spin-impurity dynamics in one-dimensional Bose-Hubbard systems~\cite{Fukuhara:2013} and ion chains~\cite{Jurcevic:2014}. In the case of Hubbard systems, the spin-description is precise only for a single spin impurity in the deep Mott insulating limit at zero temperature, where on-site number fluctuations are strongly suppressed. We will come back to this point in more detail in Sec.~\ref{case} and first neglect on-site number fluctuations. For the ion trap implementation, the correct description would be a long-range XX-model instead of the nearest-neighbor Hamiltonian~\eqref{heisenberg}. Nonetheless, the following discussion still applies to this case with a simple substitution as detailed below.\\ 
In the following, we will write a state with a single up-spin impurity on site $j$ as 
\begin{align}
\ket{j}\defs\ket{\downarrow_{-L/2},...,\downarrow_{j-1},\uparrow_j,\downarrow_{j+1},..,\downarrow_{L/2-1}},\nonumber
\end{align}\\
where $L$ is the total number of sites, and $\ket{\uparrow}$ ($\ket{\downarrow}$) refers to up-spin (down-spin) states in the z-basis. As an initial state, we choose a single up-spin impurity at the center of the chain $\ket{\psi_0}=\ket{j=0}$. For an infinite chain ($L\rightarrow \infty$), the time-evolution under Hamiltonian (\ref{heisenberg}) leads to a spreading of this impurity according to
\begin{align}
\ket{\psi_0}(t)&=  \sum_j \phi_j \ket{j}, \label{single_spin_state}
\end{align}
with $\phi_j=i^{j} J_{j}(J_{\rm ex} t/ \hbar)$, where $J_{j}(x)$ is the Bessel function of the first kind, $t$ is the evolution time, and $\hbar$ is the reduced Planck constant. For the long-range XX-model, which is relevant for ion chains, $\phi_j$ must be substituted by a different function that can be calculated numerically~\cite{Jurcevic:2014}.\\
For the experimental observation in a Hubbard system~\cite{Fukuhara:2013}, the probability of finding the spin impurity on site $j$ after various evolution times was observed in quantitative agreement with Eq.~(\ref{single_spin_state}). However, this experiment did not quantify the correlations and entanglement between spins on different sites $A$ and $B$. This information is encoded in the two-site reduced density operator $\hat \rho^s_{A,B}(t)=\text{Tr}_{l\neq A,B}[\ket{\psi_0}(t)\bra{\psi_0}(t)]$, where the trace runs over all sites but $A$ and $B$. The superscript $s$ stands for single spin-impurity. We find
\begin{align}
\hat \rho^s_{A,B}(t)&=\begin{pmatrix}
 0 & 0 & 0 &  0\\
 0 &  |\phi_A|^2 & \phi_A\phi_B^*  & 0\\
 0 & \phi_A^*\phi_B& |\phi_B|^2 & 0\\
  0& 0 & 0 & 1-|\phi_A|^2-|\phi_B|^2\\
\end{pmatrix}\label{impurity_matrix}
\end{align}
writing the two-site density matrix using basis states\newline $\ket{\uparrow,\uparrow},\ket{\uparrow,\downarrow},\ket{\downarrow,\uparrow},\ket{\downarrow,\downarrow}$ for the $A$ and $B$ sites. For any state with a single impurity in an otherwise polarized background, the reduced two-site density matrix has the structural form of $\hat \rho^s_{A,B}(t)$.\\
The entanglement between sites $A$ and $B$ can be quantified with the concurrence $C$~\cite{Wooters:1998}, a commonly used bi-partite entanglement measure~\cite{Mintert:2005,Horodecki:2009}. The concurrence for a general bipartite pure state $\ket{\psi_{1,2}}$ in a tensor product ${\cal H}_1 \otimes {\cal H}_2$ of two finite-dimensional Hilbert spaces ${\cal H}_1,{\cal H}_2$ can be defined as~\cite{Rungta:2001,Mintert:2004}
\begin{align}
C(\ket{\psi_{1,2}})=\sqrt{2(\avg{\psi_{1,2}|\psi_{1,2}}-\text{Tr}(\hat\rho_1^2))},
\end{align}
where $\hat \rho_1=\text{Tr}_{2}(\ket{\psi_{1,2}}\bra{\psi_{1,2}})$ is the reduced density operator of subsystem $1$. The concurrence defined in this way can also be applied to subnormalized states.\\
The concurrence $C(\hat \rho_{1,2})$ of a bipartite mixed state $\hat \rho_{1,2}$ is defined via a convex roof construction~\cite{Mintert:2004} using the infimum
\begin{align}
C(\hat \rho_{1,2})={\rm inf}\sum_i p_i C(\ket{\phi_i})\label{concurrence}
\end{align}
over all decompositions of $\hat \rho_{1,2}$ into pure states $\ket{\phi_i}$: $\hat \rho_{1,2}=\sum_i p_i \ket{\phi_i}\bra{\phi_i}$ with $p_i\geq 0$.
 Even if the global state $\ket{\psi_0}(t)$ is pure, the reduced density operator $\hat \rho^s_{A,B}(t)$ is mixed. We are therefore dealing with a mixed bipartite two spin-$1/2$ system.\\
Due to the X-matrix form of $\hat \rho_{A,B}^s(t)$, the concurrence can be easily calculated \cite{Yu:2007} (see Eq.~(\ref{bound_x_matrix})): 
\begin{align}
C(\hat \rho_{A,B}^s(t))=2|\phi_A\phi_B^*|\nonumber,
\end{align}
a result obtained earlier in Refs.~\cite{Subrahmanyam:2004, Amico:2004}.\\
To get a better intuition for this outcome, we can restrict ourselves to sites with $A=-B$. In this case, the two-site density matrix can be written as a mixture of a Bell-state $\ket{\Psi^+}=\frac{1}{\sqrt{2}}(\ket{\uparrow,\downarrow}+\ket{\downarrow,\uparrow})$ and $\ket{\downarrow,\downarrow}$:
\begin{align}
\hat \rho_{A,-A}^s(t)=&2 |\phi_A|^2 \ket{\Psi^+}\bra{\Psi^+}\nonumber\\
&+(1-2|\phi_A|^2)\ket{\down,\down}\bra{\down,\down}.\nonumber
\end{align}
Therefore, the concurrence amounts to the probability of finding the system in the Bell state.\\
We show $C(\hat \rho^s_{A,B}(t))$ for various times and sites $A$ and $B$
in Fig.~\ref{fig:Figure1}a and b, which illustrates how entanglement is generated and spreads in a wave-like fashion during the impurity dynamics.
\section{Scheme for spin-$1/2$ systems}\label{spinresult}
Experimentally, we are facing the problem of detecting the concurrence of an unknown two-site density matrix $\hat \rho_{A,B}$ that might be close to but not necessarily equal to $\hat \rho^s_{A,B}$ due to experimental imperfections. Detecting the concurrence of an unknown state is possible using a full state tomography. For two spin-$1/2$ systems, a full state tomography can be achieved by measuring all nine combinations of Pauli operators $\langle\hat \sigma_A^\alpha \hat \sigma_B^\beta \rangle$ with $\alpha,\beta=x,y$ and $z$~\cite{Roos:2004}. We assume that the final measurement is always performed in the $z$-basis, for example, by reading out the populations of two atomic energy levels that encode the spin-$1/2$ system. \\
A measurement in a different basis is possible by applying pulses that rotate the individual spins before the measurement. A pulse on a single spin on site $j$ can be represented with a unitary operator
\begin{align}
\hat R(\theta,\phi)_j=\begin{pmatrix}
\cos(\theta/2) & i e^{i\phi}\sin(\theta/2) \\
ie^{-i\phi}\sin(\theta/2) & \cos(\theta/2) \\
\end{pmatrix}\label{rotation_operator_12}
\end{align}
written in the $\ket{\uparrow},\ket{\downarrow}$ basis. For example, a measurement in the $x$ basis can be realized by a $\theta=\pi/2, \phi=-\pi/2$ pulse because $\hat \sigma_j^x=\hat R\left(\pi/2,-\pi/2\right)_j^\dagger\hat \sigma_j^z\hat R\left(\pi/2,-\pi/2\right)_j$.\\ 
For the following discussion, it is important to distinguish pulses on individual spins, which allow for the measurement of $\langle\hat \sigma_A^\alpha \hat \sigma_B^\beta \rangle$ for all combinations $\alpha,\beta=x,y$ and $z$, and global pulses on both spins, which restrict the measurements to equal axes $\alpha =\beta$. \\
Pulses on individual spins arranged in a chain are commonly employed in ion trap implementations. For example, in Ref.~\cite{Jurcevic:2014}, the authors show the detection of the concurrence generated during spin impurity dynamics in a long-range XX-model using a full state tomography.\\ 
However, for optical lattice implementations of spin-systems using Hubbard models, only $\theta=\pi$ pulses on individual atoms have been demonstrated using a rapid adiabatic passage~\cite{Weitenberg:2011}. Pulses on individual atoms with arbitrary $\theta,\phi$ require improved experimental control and are yet to be implemented. This currently restricts the detection to elements $\langle\hat \sigma_A^\alpha \hat \sigma_B^\beta \rangle$ with $\alpha = \beta$. Therefore, we now present a simplified scheme for the detection of a lower bound for the concurrence using only global pulses.
\subsection{Bound for global pulses with controlled $\phi$}
The first step in deriving the bound is to split the unknown two-site density matrix into an X- and O-part according to
\begin{align}
\hat \rho_{A,B}&=\hat X+\hat O \nonumber
\end{align}
with
\begin{align}
\hat X\defs\begin{pmatrix}
 P_{\up,\up} & 0 & 0 &  \rho_{\up \up}\\
 0 &  P_{\up,\down} & \rho_{\up\down} & 0\\
 0 & \rho_{\up\down}^* & P_{\down,\up} & 0\\
 \rho_{\up \up}^* & 0 & 0 & P_{\down,\down}\\
\end{pmatrix}\nonumber
\end{align}
and $\hat O=\hat \rho_{A,B}-\hat X$.\\
Knowledge of only the X-part is sufficient to detect a lower bound for the concurrence because~\cite{Jafarpour:2011,Rafsanjani:2012}:
\begin{align}
C(\hat X)\leq C(\hat \rho_{A,B}).\nonumber
\end{align} 
The concurrence for density matrices in X-form is given by \cite{Yu:2007}
\begin{align}\label{bound_x_matrix}
C(\hat X)=2 {\rm max}(0,|\rho_{\up\up}|-\sqrt{P_{\up,\down}P_{\down,\up}},|\rho_{\up\down}|-\sqrt{P_{\up,\up}P_{\down,\down}}).
\end{align}
Because
\begin{align}
\rho_{\up\down}&=\frac{1}{4}(\langle \hat \sigma_A^x \hat\sigma_B^x \rangle+\langle \hat \sigma_A^y \hat \sigma_B^y \rangle+i(\langle \hat \sigma_A^x \hat \sigma_B^y \rangle+\langle \hat\sigma_A^y \hat\sigma_B^x \rangle)),\nonumber
\end{align}
we have $\frac{1}{4}|\langle \hat \sigma_A^x \hat\sigma_B^x\rangle+\langle\hat \sigma_A^y\hat \sigma_B^y\rangle|\leq |\rho_{\up\down}|$. Thus, we find the lower bound
\begin{align}
2\left(\frac{1}{4}|\langle\hat \sigma_A^x \hat \sigma_B^x\rangle+\langle\hat \sigma_A^y \hat \sigma_B^y\rangle|-\sqrt{P_{\up,\up} P_{\down,\down}}\right)\leq C(\hat \rho_{A,B}) \label{bound1},
\end{align}
which only requires global pulses for the detection of $\langle\hat \sigma_A^x \hat \sigma_B^x\rangle$ and $\langle\hat \sigma_A^y \hat \sigma_B^y\rangle$. The probabilities for having both spins up, $P_{\up,\up}$, and both spins down, $P_{\down,\down}$, can be detected in the z-basis without pulse before the measurement. 
\subsection{Bound for global pulses with undetermined phase $\phi$}\label{no_phase}
The phase $\phi$ of the applied pulse is difficult to control experimentally. For the case of the impurity dynamics detailed above, controlling the phase $\phi$ would require having a defined phase of the applied field for the pulse relative to the starting time of the dynamics. This is difficult to reach for the implementation in Ref.~\cite{Fukuhara:2013} because the spin dynamics occurs in the tens of hertz regime, while the applied pulses are in the gigahertz regime. We will assume that the pulses are not phase-locked to the starting point of the dynamics. In this case, $\phi$ is essentially random. All observables after a global pulse with $\theta$ are then effectively described by an equal statistical mixture over all angles $\phi$ described by a density matrix
\begin{align}
\hat \rho_{A,B}(\theta)=\frac{1}{2\pi}\int_0^{2\pi}\!\!\!d\phi\,\hat \rho_{A,B}(\theta,\phi)\nonumber,
\end{align}
where 
\begin{align}
\hat \rho_{A,B}(\theta,\phi)=\hat R(\theta,\phi)_A \hat R(\theta,\phi)_B \hat \rho_{A,B} \hat R(\theta,\phi)_B^\dagger \hat R(\theta,\phi)_A^\dagger \nonumber
\end{align}
is the two-site density matrix after a global pulse with angles $\theta$ and $\phi$.\\
Let us denote the average value of $ \hat \sigma_A^z \hat \sigma_B^z $ after a global pulse with $\theta=\pi/2$ and random $\phi$ by $\langle\hat \sigma_A^z \hat \sigma_B^z \rangle_{\pi/2}$. Then, we have
\begin{align}
\langle \sigma^z_A \sigma^z_B \rangle_{\pi/2}&\defs{\rm Tr}[ \hat \rho_{A,B}(\pi/2)\hat \sigma^z_A\hat \sigma^z_B ]\nonumber\\
&=\frac{1}{2}\left(\langle \hat\sigma^x_A \hat \sigma^x_B \rangle+\langle\hat \sigma^y_A \hat \sigma^y_B\rangle\right)\nonumber,
\end{align}
where we used the invariance of the trace under cyclic permutation in the second line. Therefore, a measurement after a global $\pi/2$ pulse with random $\phi$ corresponds to a measurement of the mean of $\langle \hat \sigma^x_A \hat \sigma^x_B \rangle$ and $\langle \hat \sigma^y_A \hat \sigma^y_B \rangle$.\\
The bound (\ref{bound1}) can then be rewritten as
\begin{align}
K(\hat \rho_{A,B})\defs 2\left(\frac{1}{2}|\langle \hat \sigma^z_A \hat \sigma^z_B \rangle_{\pi/2}|-\sqrt{P_{\up,\up} P_{\down,\down}}\right)\leq C(\hat \rho_{A,B}). \label{bound2}
\end{align}
Importantly, a detection of $K_{A,B}$ only requires measurements with and without a global $\pi/2$ pulse (with random phase $\phi$), which simplifies the experimental effort dramatically as compared to a full state reconstruction.
\subsection{Quality of the bound for the case of single-spin dynamics}\label{quality}
An important question is how tight the bound (\ref{bound2}) is for the case of single-impurity dynamics detailed in Sec.~\ref{single_impurity}. Using Eq.~(\ref{impurity_matrix}), we find that
\begin{align}
K(\hat \rho^s_{A,B})=  \begin{cases}
   C(\hat \rho^s_{A,B}) & \text{if } A-B\,\,\, \rm{even} \\
   0       & \text{if } A-B \,\,\, \rm{odd}.
  \end{cases}\nonumber
\end{align}
The reason for this behavior is that spins at even distances have a parallel alignment in the $x-y$ plane in the sense that \mbox{$|\langle \hat\sigma^x_A \hat \sigma^x_B\rangle|=|\langle \hat \sigma^y_A \hat \sigma^y_B\rangle|>0$} and $|\langle \hat \sigma^x_A \hat \sigma^y_B\rangle|=|\langle \hat \sigma^x_A \hat \sigma^y_B\rangle|=0$. In contrast, for odd distances, the spins have a perpendicular alignment, $|\langle \hat\sigma^x_A  \hat \sigma^x_B\rangle|=|\langle\hat \sigma^y_A  \hat \sigma^y_B\rangle|=0$ and $|\langle\hat \sigma^x_A  \hat \sigma^y_B\rangle|=|\langle \hat \sigma^y_A \hat\sigma^x_B\rangle|>0$. \\
This even-odd behavior leads to a peculiar checkerboard pattern if $K(\hat\rho^s_{A,B})$ is plotted as a function of $A$ and $B$ (Fig.~\ref{fig:Figure1}c). While the fact that $K(\hat \rho^s_{A,B})=0$ for odd distances is a disadvantage on first glance, this checkerboard pattern can serve as an experimental signature on top of noisy experimental data.\\
Without going into details, we note that by applying a magnetic field gradient before the detection, the off-diagonal element $\rho_{\up\down}$ acquires a time-dependent complex phase-factor. Tuning this phase to $\pi/2$ changes the parallel alignment of the spins into perpendicular alignment and vice versa. As a result, the measured bound would be tight for odd distances and zero for even distances. Using this technique, a tight bound can be achieved for all pairs of spins.
\section{Spin-entanglement in the presence of atom number fluctuations}\label{number_flctuations}
Quantum magnetism experiments in optical lattices are typically performed using mixtures of atoms in two different hyperfine states~\cite{Kuklov:2003, Duan:2003, Trotzky:2008a, Fukuhara:2013, Fukuhara:2013b,Greif:2013,Nascimbene:2013}. The local on-site states can be written as $\ket{n_{l}^+,n_{l}^-}$, where $n_{l}^+$ and $n_{l}^-$ is the number of atoms in the two hyperfine states on site $l$. The state of the whole system can be expanded in basis states $\prod_l \ket{n_{l}^+,n_{l}^-}$. We also introduce a notation for the total atom number on site $l$ as $n_l=n_l^++\,n_l^-$.
The connection to spin systems is obtained using the Schwinger representation (see, e.g., Ref. \cite{sakurai:1993}), which maps the on-site states to a total spin $j_l$ system with spin-projection $m_l$ defined as
\begin{align}
j_l&=\frac{n_l^++n_l^-}{2},\,\, m_l=\frac{n_l^+-n_l^-}{2}\nonumber.
\end{align}
We will also use the notation $\ket{j_l,m_l}=\ket{n_{l}^+,n_{l}^-}$.\\
In the large-interaction limit of Hubbard models, the dynamics in subsectors with fixed $j_l=1/2$ is governed by XXZ-models~\cite{Kuklov:2003, Duan:2003}. However, due to the finite temperature of the samples~\cite{Bakr:2010,Endres:2011} and quantum fluctuations~\cite{Sherson:2010}, number fluctuations are introduced into the system. This results in contributions of on-site states that map to different $j_l\neq 1/2$.\\
We are facing a situation where both spin fluctuations (i.e., fluctuations of $m_l$ for a fixed $j_l$) and number fluctuations (i.e., fluctuations of $j_l$) are present in the system. It is both experimentally and conceptually interesting to ask whether entanglement between the spin-projection degree of freedom is detectable in this scenario. \\
Again, we consider a subsystem consisting of two sites $A$ and $B$, for which the reduced density operator now also includes contributions from different occupation numbers:
\begin{align}
&\hat \rho_{A,B}\nonumber\\
&=\sum_{j_A,j'_A,...}\rho_{\s{j'_A,m'_A,j'_B,m'_B\\j_A,m_A,j_B,m_B}}\ket{j'_A,m'_A,j'_B,m'_B}\bra{j_A,m_A,j_B,m_B}ß\nonumber\\
&=\sum_{n^-_A,\bar n^+_A,...}\rho_{\s{\bar n^+_A,\bar n^-_A,\bar n^+_B,\bar n^-_B\\n^+_A,n^-_A, n^+_B,n^-_B}}\ket{\bar n^+_A,\bar n^-_A,\bar n^+_B,\bar n^-_B}\bra{n^+_A,n^-_A, n^+_B,n^-_B}\nonumber,
\end{align}
where we used the Schwinger and occupation number notation in the second and last line respectively.
\subsection{Entanglement of particles}
First, we are dealing with the question of how to conceptually differentiate  the entanglement in the spin degree of freedom from entanglement that stems from different total local occupation numbers~\cite{Wiseman:2003,Dowling:2006,Schuch:2004}. For example, superpositions of states with different local atom numbers, such as $\frac{1}{\sqrt{2}}(\ket{1,0,0,0}+\ket{0,0,1,0})$ (corresponding to a single plus atom in a superposition between site $A$ and $B$), should not appear entangled.\\
An appropriate procedure to achieve this goal is to first project onto states with fixed local atom numbers. To this end, we define  projected two site operators
\begin{align}
\hat{\rho}^{n_A,n_B}_{A,B}=\hat\Pi^{n_A}_A\hat\Pi^{n_B}_B \hat \rho_{A,B}\hat\Pi^{n_A}_A\hat\Pi^{n_B}_B,\label{number_projected}
\end{align}
where
\begin{align}
&\hat\Pi^{n_l}_l =\sum_{m_l}\ket{j_l=n_l/2, m_l}\bra{j_l=n_l/2, m_l}\nonumber
\end{align}
is the projection operator at site $l$ onto local total atom number $n_l=n_l^++n_l^-$, or, in the Schwinger notation, onto local total spin $j_l=n_l/2$.\\
The entanglement in the spin degree of freedom can then be captured by the so-called entanglement of particles~\cite{Wiseman:2003,Dowling:2006}
\begin{align}
E_p(\hat \rho_{A,B})&=\sum_{n_A,n_B} p^{n_A,n_B} C(\hat\rho_{A,B}^{n_A,n_B}/p^{n_A,n_B})\nonumber\\
&=\sum_{n_A,n_B}  C(\hat\rho_{A,B}^{n_A,n_B})\label{eparticle},
\end{align}
where we used the concurrence $C$ as an entanglement measure, and 
\begin{align}
p^{n_A,n_A}\defs{\rm Tr}[\hat{\rho}^{n_A,n_B}_{A,B}] \nonumber
\end{align}
is the probability of finding the system with $n_A$ atoms on $A$ and $n_B$ atoms on $B$.\\
In the first line of \eqref{eparticle}, the concurrence is evaluated with the normalized state $\hat\rho_{A,B}^{n_A,n_B}/p^{n_A,n_B}$. The second line follows from the definition \eqref{concurrence} of the concurrence applied to the subnormalized operator $\hat\rho_{A,B}^{n_A,n_B}$.\\
A trivial lower bound for $E_p(\hat \rho_{A,B})$ is 
\begin{align}
C(\hat\rho_{A,B}^{1,1})\leq E_p(\hat\rho_{A,B}). \label{trivial_bound}
\end{align}
The projected operator $\hat\rho_{A,B}^{1,1}$ describes the subsector with unity filling on both sites, that is, with local total spin $j_l=1/2$ on both sites. We will refer to this as the \mbox{spin-$1/2$} sector in the following.\\
Our goal is to formulate a detectable lower bound for the entanglement contained in the spin-1/2 sector quantified by the concurrence $C(\hat\rho_{A,B}^{1,1})$, which can eventually be used to bound the entanglement of particles via the previous inequality.
\subsection{Simplified spin-1/2 notation}
For the following sections, we will introduce a shorthand notation for the density matrix elements in the spin-1/2 sector based on the Schwinger notation:
\begin{align}
\rho_{\s{m_A,m_B\\m'_A,m'_B}}=\rho^{1,1}_{\s{j_A=1/2,m_A,j_B=1/2,m_B\\j'_A=1/2,m'_A,j'_B=1/2,m'_B}}\nonumber.
\end{align}
Instead of the cumbersome notation with $\pm\frac{1}{2}$, we will use $\uparrow$ and $\downarrow$ for spin up and spin down. For example, 
\begin{align}
\rho_{\up\down}=\rho_{\s{m_A=1/2,m_B=-1/2\\m'_A=-1/2,m'_B=1/2}}\nonumber
\end{align}
and
\begin{align}
\ket{\uparrow,\downarrow}=\ket{j_A=\frac{1}{2},m_A=\frac{1}{2},j_B=\frac{1}{2},m_B=-\frac{1}{2}}.
\end{align}
\section{Case study: Entanglement during spin-impurity dynamics in the Bose-Hubbard model}\label{case}
To illustrate these concepts and to investigate the influence of number fluctuations, we carried out a case study by numerically simulating the dynamics of a mobile spin impurity in the one-dimensional two-species Bose-Hubbard model:
\begin{align}
 \hat H_{\rm BH} =& -J \sum_{\sigma, j} \left( \hat b_{\sigma,j}^\dagger \hat b_{\sigma, j+1} + H.c. \right) + \nonumber \\
 &+ \frac{U}{2} \sum_{\sigma, \sigma', j} \hat n_{\sigma,j} (\hat n_{\sigma',j}-\delta_{\sigma, \sigma'}).
 \label{eq:BoseHubbard}
\end{align}
Here $\hat b^{(\dagger)}_{\sigma,j}$ is the operator that annihilates (creates) a boson of species $\sigma = \{+,-\}$ at site $j$, $J$ is the hopping amplitude and $U$ the interaction strength. Note that for simplicity the inter- and intra-species interaction parameters are taken to be equal, although in usual alkaline gases they assume slightly different values.\\
In the limit $U \gg J$ at $N^++N^- = L$ ($N^\pm$ is the total atom number of the respective species), the system is in a Mott phase with one particle per site, where charge degrees of freedom are frozen, but internal ones are not. 
They can be described with an XXZ Hamiltonian via second-order perturbation theory~\cite{Kuklov:2003, Duan:2003}:
\begin{equation}
 \hat H_{\rm XXZ} = - \frac {J_{\rm ex}}2 \sum_j \left( \hat S_j^+ \hat S_{j+1}^- + \hat S_j^- \hat S_{j+1}^+\right) -
J_{\rm ex} \sum_j \hat S_j^z \hat S_{j+1}^z,
 \label{eq:XXZ}
\end{equation}
where $J_{\rm ex} = 4 J^2 /U$.
The local states $| \uparrow \rangle$ and $| \downarrow \rangle$
upon which the spin-1/2 operators $\hat S_j^\pm$ act
are identified with the states $|n^+ = 0, n^- = 1 \rangle$ and $|n^+ = 1, n^- = 0 \rangle$, respectively, using the Schwinger representation (see  Sec.~\ref{number_flctuations}). For the case of a single spin impurity in an otherwise polarized chain, the last term of $\hat H_{\rm XXZ}$ is only a constant offset, and the dynamics are described by the XX Hamiltonian $\hat H_{\rm XX}$ as discussed in Sec.~\ref{single_impurity}.\\
Due to on-site number fluctuations, this mapping can break down in experimentally relevant parameter ranges. We consider two possibilities in the following. First, for stronger hopping $J$, significant quantum fluctuations of the on-site particle number are introduced in the form of correlated particle-hole pairs~\cite{Endres:2011} even at zero temperature. One of the open questions here is up to which dimensionless hopping strength $J/U$ the spin description holds. Second, at finite temperature, thermally excited defects can lead to a break down of the spin-description even for values of $J/U$ where the XXZ model would be a very good approximation at zero temperature. In this case, a crucial question concerns the temperature range in which an observation of spin-entanglement is experimentally feasible. 
\begin{figure}[t]
\begin{center}
 \includegraphics[width=\columnwidth]{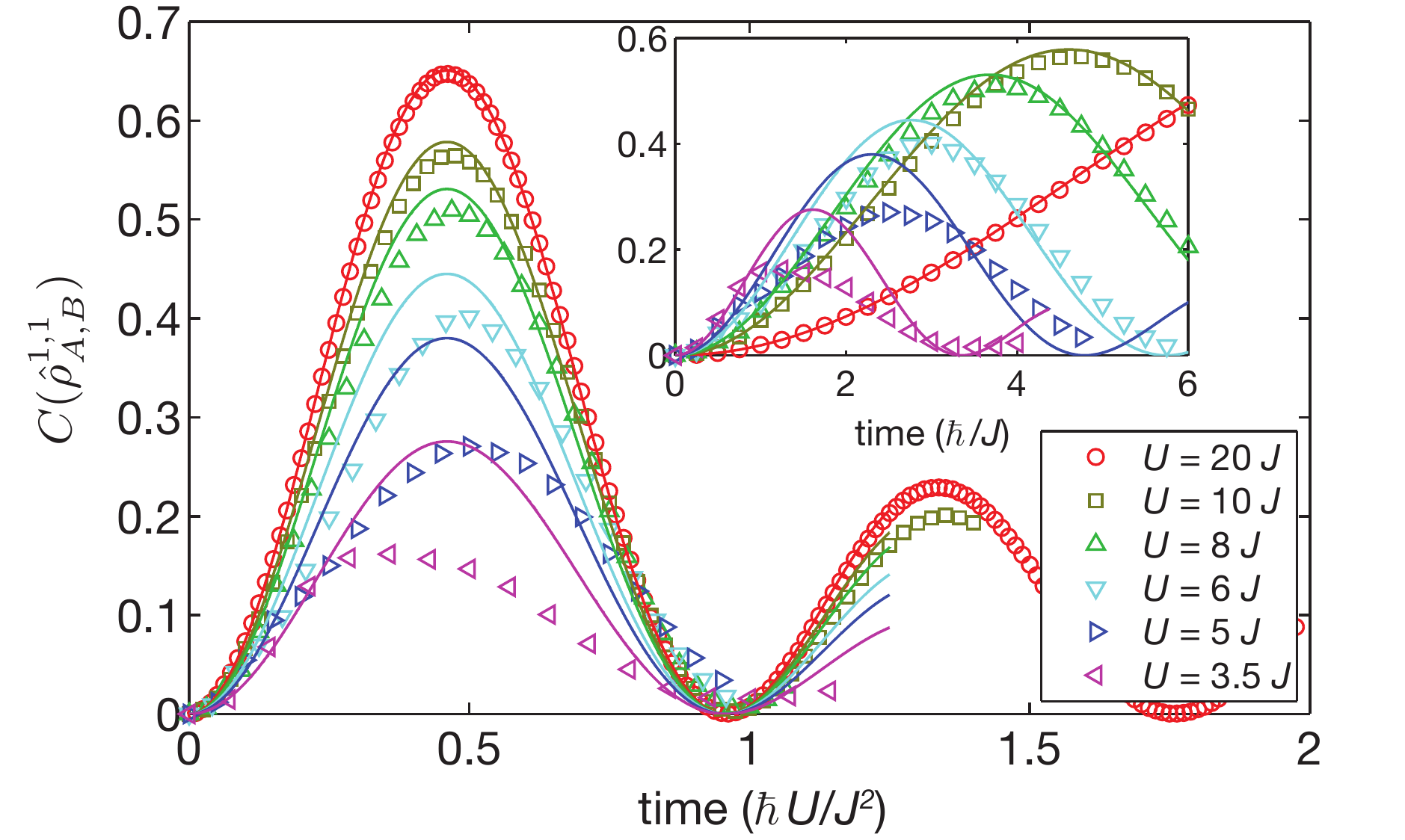}
\end{center}
\caption{Concurrence $C(\hat\rho_{A,B}^{1,1})$ for the subsector with a single particle per site for the sites $A=L/2+1$ and $B=L/2-1$ as a function of time for several values of $U/J$ computed with TEBD (for markers and colors, see legend). The x-axis is rescaled by the typical spin exchange coupling $J^2/U$. The prediction for the XX model from Eq.~\eqref{single_spin_state} rescaled by $p^{1,1}(t=0)$ is shown as solid lines in the same colors as the respective numerical data. Inset: same data as main plot with x-axis rescaled by the hopping $J$.}
\label{fig:1}
\end{figure}
\subsection{Influence of Quantum Fluctuations}
To investigate the influence of quantum fluctuations, we studied the situation $U \gtrsim J$, where the system is in a Mott insulating phase but particle fluctuations are not negligible~\cite{Endres:2011} using algorithms based on Matrix-Product-States~\cite{Schollwoeck:2011}. A system of size $L=30$ is initialized in the ground state of Hamiltonian~\eqref{eq:BoseHubbard} in the sector where $N^- = L$ and $N^+ = 0$. Since we consider the regime $20 \geq U/J \geq 3.5$,
this correspond to a Mott insulating phase of the $\sigma = -$ bosons in the thermodynamic limit~\cite{Kuehner:1998}.\\
We subsequently perform a spin flip for the central spin of the chain using the protocol: \mbox{$|n^+=0, n^- = 0 \rangle_{L/2} \to |0, 0 \rangle_{L/2}$} and
\mbox{$|0, n^- \rangle_{L/2} \to |1, n^--1 \rangle_{L/2}$}. With this protocol, we need to consider a local Hilbert space that has to accommodate at most one $\sigma = +$ boson per site, simplifying the numerical simulation. For the $\sigma =-$ bosons, we truncate their local Hilbert space to four occupancies,
with the further constraint that there can be 
at most four particles per site (the state $|1,4 \rangle_{L/2}$
is thus discarded). \\
The system is then evolved in time with Hamiltonian~\eqref{eq:BoseHubbard} using a time-evolving block decimation algorithm (TEBD)~\cite{Vidal:2003}. During the time-evolution the maximal allowed bond link is $D=3000$.\\
In Fig.~\ref{fig:1}, we show the concurrence $C(\hat\rho_{A,B}^{1,1})$ for the subsector with a single particle per site for the sites $A=L/2+1$ and $B=L/2-1$ as a function of time for several values of $U/J$.
Oscillations have a clear $U/J^2$ period, which is the time-scale
associated with the typical energy scale of spin dynamics $J_{\rm ex}$. A clear decrease of the maximum concurrence for lower $U$ is visible.\\
One reason for this decrease is that for lower $U$, the probability $p^{1,1}$ to find a single particle per site is reduced, which corresponds to a reduced trace of $\hat\rho_{A,B}^{1,1}$. To check for this effect, we compare the dynamics to the prediction from Eq.~\ref{single_spin_state} for the XX model rescaled by $p^{1,1}(t=0)$. The curves for the rescaled XX dynamics are shown in Fig.~\ref{fig:1} as solid lines. For $U/J \gtrsim 8 $, the dynamics appears to be well described by the rescaled XX predictions, indicating that effective spin dynamics in the sector with one particle per site are undisturbed by the presence of number fluctuations. For lower $U/J$, the concurrence $C(\hat\rho_{A,B}^{1,1})$ is smaller than predicted by the rescaled solution.\\
\begin{figure}[t]
\begin{center}
\includegraphics[width=\columnwidth]{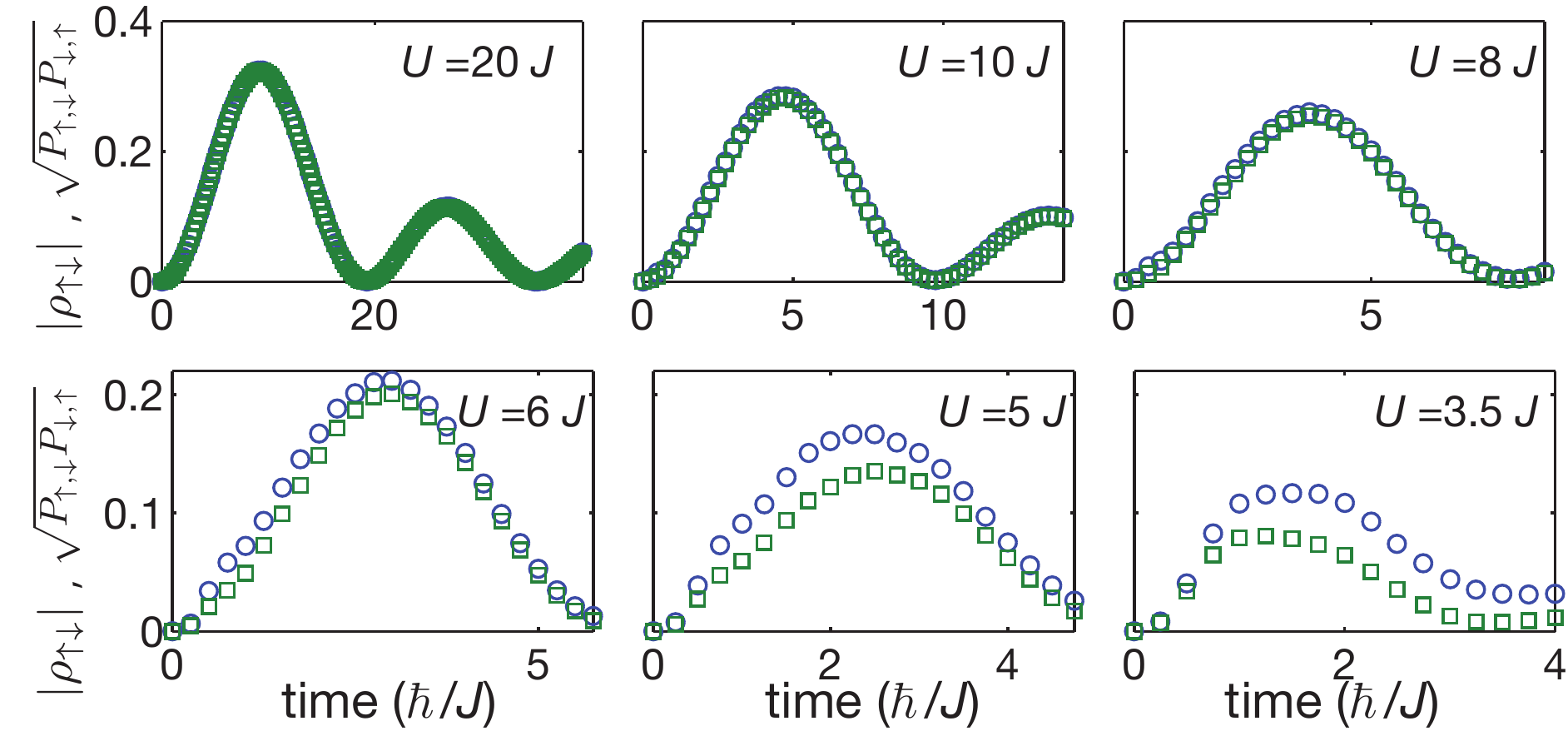}
\end{center}
\caption{Time evolution of 
$\sqrt{P_{\uparrow, \downarrow} P_{\downarrow, \uparrow}}$ 
(blue circles) and  
$|\rho_{\uparrow \downarrow }|$
(green squares) for different values of $U/J$ computed with TEBD. 
For lower $U/J$, the two quantities start to deviate, signaling a decoherence process due to quantum fluctuations.}
\label{fig:2}
\end{figure}
We now inspect the reduced density operator $\hat \rho^{1,1}_{A,B}$ more closely by comparing the quantities 
$\sqrt{P_{\uparrow, \downarrow} P_{\downarrow, \uparrow}}$
and 
$|\rho_{\uparrow \downarrow }|$,
which are equal in the spin case [see Eq.~\eqref{impurity_matrix}]. The equality of both quantities signals fully coherent dynamics. In Fig.~\ref{fig:2}, we show the time evolution of both quantities for several values of $U/J$. Interestingly, the two quantities take similar values down to $U/J \sim 6$. The fact that we observe $|\rho_{\uparrow \downarrow }|
<
\sqrt{P_{\uparrow, \downarrow} P_{\downarrow, \uparrow}}$ for lower values of $U/J$
can be interpreted as effective decoherence dynamics.\\
\begin{figure}[t]
\begin{center}
\includegraphics[width=\columnwidth]{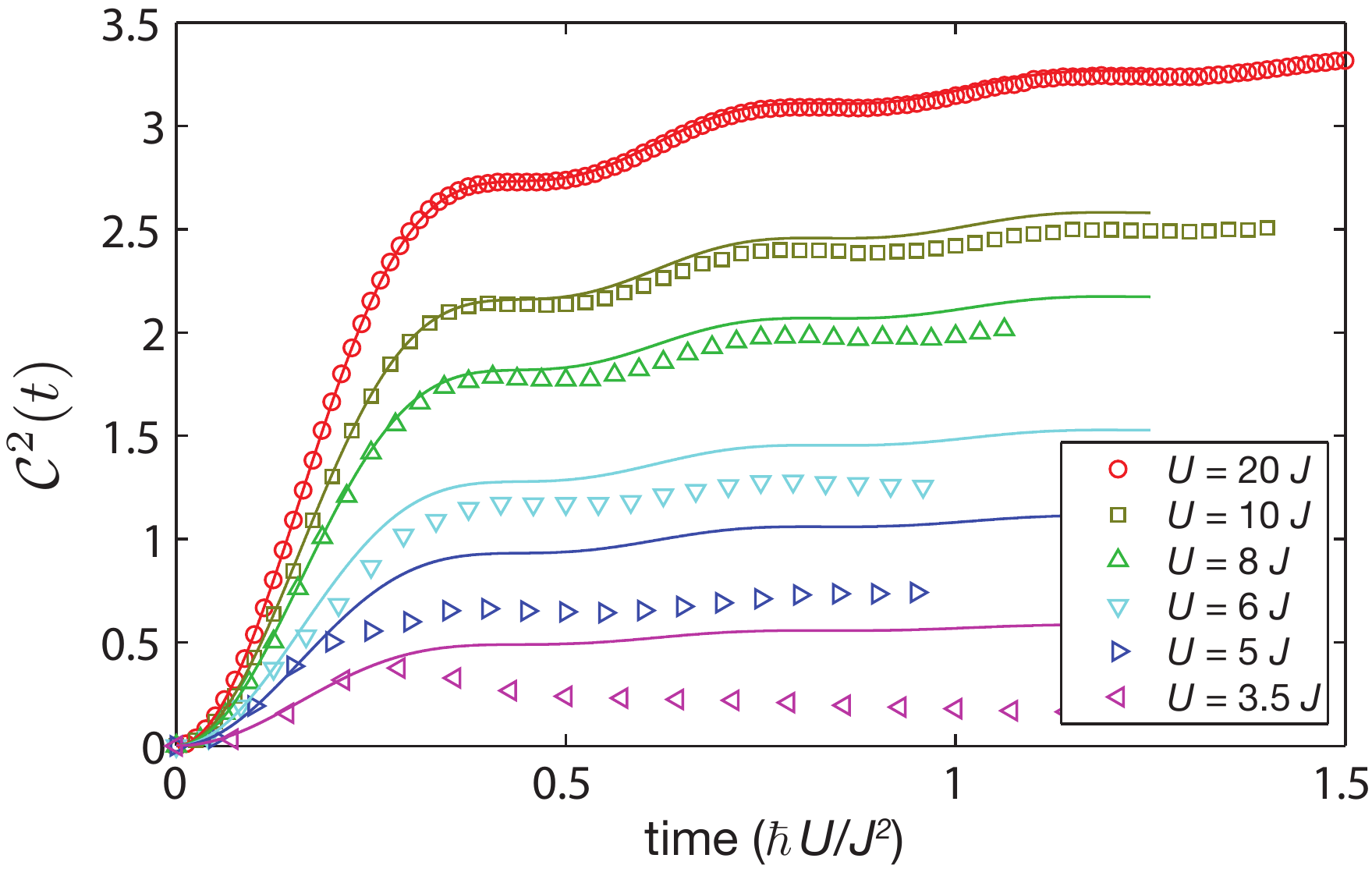}
\end{center}
\caption{Global measure of entanglement, ${\cal C}^2(t)$,
for several values of $U/J$. Solid lines show the rescaled XX prediction.}
\label{fig:4}
\end{figure}
Finally, we consider a global measure of entanglement in the system by investigating the sum of the squared concurrences:
\begin{equation}
{\cal C}^2(t) \defs \sum_{i,j} C^2(\hat \rho^{1,1}_{i,j} (t))\nonumber.
\end{equation}
The motivation for summing over the square of the concurrences stems from the monogamy inequality~\cite{Osborne:2006}, which holds for spin-$1/2$ systems. For the ideal spin dynamics in the XX-Hamiltonian, \mbox{${\cal C}^2(t)=4(1-\sum_A |\phi_A(t)|^2)\rightarrow 4$} for long times.\\
In Fig.~\ref{fig:4}, we show $\mathcal C^2(t)$ for several values of $U/J$. For $U/J \gtrsim 8$, $\mathcal C^2(t)$ increases with time, and the prediction of the XX chain weighted with the probability $p^{1,1}(t=0)$ (solid lines) captures the behavior. For smaller $U/J$, stronger deviations are visible, which indicates decoherence.\\
\subsection{Influence of Thermal Fluctuations}
\begin{figure*}[t]
\begin{center}
\includegraphics[width=\textwidth]{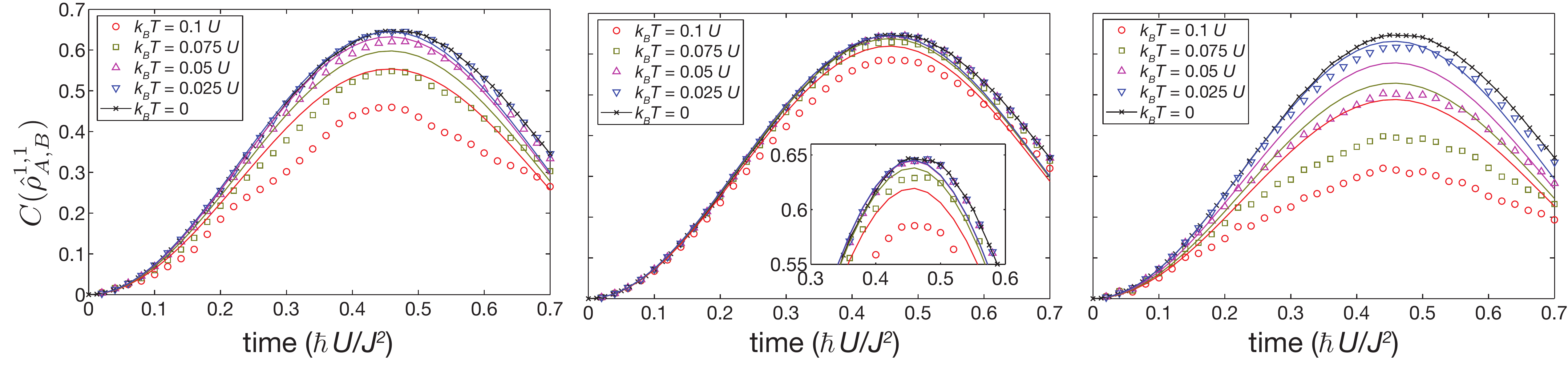} 
\end{center}
\caption{Concurrence $C(\hat\rho_{A,B}^{1,1})$ for the subsector with a single particle per site for the sites $A=L/2+1$ and $B=L/2-1$ as a function of time for several values of inverse temperature $\beta$ (for markers and colors, see legend) and chemical potentials
$\mu = U/4$, $U/2$ and $3U/4$ (from left to right panel). Thin solid lines represent the prediction of the XX spin model rescaled 
by $p^{1,1}(t=0)$.
}
\label{fig:7}
\end{figure*}
We now turn to the effects of number fluctuations introduced by a finite temperature and a chemical potential.
In order to single out these effects, the hopping strength is set to $J=20U$, for which the dynamics at zero temperature is well captured by the XX prediction as shown in the previous section. To this end, we performed exact diagonalization of a system of size $L=6$
and included only the lowest-energy part of the Hilbert space. As the hopping of particles is only a small perturbation, we consider only Fock states with an interaction energy smaller than a given energy cutoff: 
$ (U/2) \langle\sum \hat n_{\sigma,j} (\hat n_{\sigma',j}-\delta_{\sigma, \sigma'})\rangle <E_c$. 
The system is initialized in the grand canonical ensemble
of $\sigma = -$ bosons with temperatures in a range $ T = 0 - 0.1 U/k_{\rm B}$ ($k_{\rm B}$, Boltzmann constant) and chemical potentials in a range $\mu = 0.25 U- 0.75 U$. We perform the same flip protocol as in the previous section and let the system 
evolve in time with Hamiltonian $\hat H_{\rm BH}$~\eqref{eq:BoseHubbard}. The simulation includes a total number of particles $N_{\rm tot}=(N^++N^-) \in [4,8]$ and the cutoff energy is $E_c=3 U + \mu (N_{\rm tot}-L)$. Convergence of the simulations upon
inclusion of more particle sectors and more states has been verified and an error on the order of a few percents is estimated, which is better than the expected experimental precision. In order to test the influence of the relatively small size of $L=6$ on 
the time-evolution, we compared the zero-temperature concurrence spreading at $L=6$ with the spreading at $L=30$ with the TEBD (see previous section). We find that until time $t \sim 0.5 \hbar U/J^2$, the two predictions agree within a few percent. Even if finite-size corrections are expected to be more significant at higher temperatures (we compared the data with those at L=5, not shown), the data in Fig. 5 should be sufficiently accurate to predict the behavior of typical experimental systems with $L\approx 15 - 20$ within a few percent.\\
In Fig.~\ref{fig:7}, we show the concurrence $C(\hat \rho^{1,1}_{L/2-1,L/2+1})$ for chemical potentials $\mu=0.25U, 0.5U, 0.75U$ and several values of the temperature $T$. For increasing temperatures the signal drops. This reduction with temperature is relatively small at $\mu=0.5U$ compared with the other two chemical potential values. This can be attributed to the fact that the gap to excited states is largest, and thermal excitations are thus suppressed, at $\mu=0.5U$ in the limit $J/U=0$~\cite{Sherson:2010}. This statement holds in approximate form also for $J/U=1/20$. Additionally, the effect of an increase of the chemical potential from the optimal value $\mu\approx 0.5 U$ to $\mu=0.75U$ is more damaging to the entanglement than a decrease to $\mu=0.25U$ (compare left and right plot in Fig.~\ref{fig:7}). This dependence on the chemical potential highlights the importance of tuning the chemical potential at the center of a trapped system to $\mu \approx 0.5U$.\\
Similar to the case of quantum fluctuations, we check whether decreased concurrence can be ascribed to the reduced population 
of the single-occupancy sector. Solid lines in Fig.~\ref{fig:7} represent the prediction of the XX model rescaled by 
$p^{1,1}(t=0)$. Whereas the XX model captures the features of the entanglement dynamics for low temperatures and for $\mu = 0.5 U$, it fails
at the highest temperatures considered.\\
Concluding, we provided evidence that the entanglement propagation scheme previously described can be carried out in a realistic parameter range for experiments. For current temperatures of $T\approx 0.1U/k_B$~\cite{Sherson:2010,Fukuhara:2013}, a drop of the concurrence signal by maximally a factor of two compared to the zero temperature situation is to be expected due to number fluctuations introduced by finite temperature in the grand canonical ensemble. Therefore, the signal should be strong enough to be experimentally detectable.
\section{Scheme in the presence of number fluctuations assuming full spin-resolution}\label{scheme_full_spin}
We now turn to the description of an entanglement detection scheme for a lower bound of the concurrence $C(\hat\rho_{A,B}^{1,1})$. In this section, we assume that the measurement can be done with full spin-resolution, that is, the individual populations $n^\pm$ of both species can be detected in a single experimental run.
\subsection{Observable with full spin-resolution}
Restricting ourselves to global pulses, the most general observable, in this case, is the joint probability of finding $n_{A}^\pm$ atoms of species $\pm$ on site $A$, and $n_{B}^\pm$ atoms of species $\pm$ on site $B$ after a global pulse with angles $\theta$ and $\phi$. This probability can be written in terms of the diagonals
\begin{align}
P_{n^+_{A},n^-_{A},n^+_{B},n^-_{B}}(\theta,\phi)\defs\rho_{\s{n^+_A, n^-_A,n^+_B, n^-_B\\n^+_A, n^-_A,n^+_B, n^-_B}}(\theta,\phi)\nonumber
\end{align}
of the reduced density operator
\begin{align}
\hat \rho_{A,B}(\theta,\phi)\defs\hat R(\theta,\phi)_A \hat R(\theta,\phi)_B \hat \rho_{A,B} \hat R(\theta,\phi)_B^\dagger \hat R(\theta,\phi)_A^\dagger , \nonumber
\end{align}
after the global pulse with angles $\theta$ and $\phi$.\\
The rotation operator $\hat R(\theta,\phi)_l$ is a generalization of the spin-$1/2$ rotation (\ref{rotation_operator_12}) to arbitrary local total spins $j_l$. It can be obtained using the transformation of the creation operators $\hat a_{l,\pm}^\dagger$ for species $\pm$ on site $l$
\begin{align}
\hat a(\theta,\phi)_{l,+}^\dagger&=\cos(\theta/2)\hat a_{l,+}^\dagger+i e^{-i\phi}\sin(\theta/2)\hat a_{l,-}^\dagger\nonumber\\
\hat a(\theta,\phi)_{l,-}^\dagger&= i e^{i\phi}\sin(\theta/2)\hat a_{l,+}^\dagger+\cos(\theta/2)\hat a_{l,-}^\dagger\nonumber,
\end{align}
which yields the mapping
\begin{align}
& \hat R(\theta,\phi)_l\ket{j_l,m_l}=\frac{(\hat a(\theta,\phi)_{l,+}^{\dagger})^{j_l+m_l}(\hat a(\theta,\phi)_{l,-}^{\dagger})^{j_l-m_l}}{\sqrt{(j+m)!}\sqrt{(j-m)!}}\ket{0}_l\label{rot_general}
\end{align}
for the basis states $\ket{j_l,m_l}$ \cite{sakurai:1993}.
\subsection{Observable for random phase $\phi$}
As discussed in Sec.\ref{no_phase}, the phase $\phi$ of the global pulse is assumed to be random. Therefore, we consider an averaged density operator
\begin{align}
\hat \rho_{A,B}(\theta)\defs\frac{1}{2\pi}\int_0^{2\pi}\!\!\!d\phi\,\hat \rho_{A,B}(\theta,\phi)\nonumber,
\end{align}
and the experimentally observed probabilities are
\begin{align}
P_{n^+_{A},n^-_{A},n^+_{B},n^-_{B}}(\theta)\defs\frac{1}{2\pi}\int_0^{2\pi}\!\!\!d\phi\, P_{n^+_{A},n^-_{A},n^+_{B},n^-_{B}}(\theta,\phi). \nonumber
\end{align}
Additionally, we are interested in the probabilities $p^{n_A,n_B}$ for observing the total atom numbers $n_A$ and $n_B$. They can be detected by summing over $P_{n^+_{A},n^-_{A},n^+_{B},n^-_{B}}(0)$ with the constraint that $n_A^++n_A^-=n_A$ and $n_B^++n_B^-=n_B$:
\begin{align}
p^{n_A,n_B}=\!\!\!\!\!\!\!\!\!\!\!\!\!\!\!\!\sum_{\s{n^+_{A},n^-_{A},n^+_{B},n^-_{B}\\ n^+_{A}+n^-_{A}=n_A,n^+_{B}+n^-_{B}=n_B}} \!\!\!\!\!\!\!\!\!\!\!\!\!\!\!\! P_{n^+_{A},n^-_{A},n^+_{B},n^-_{B}}(0). \label{pnanb}
\end{align}

\subsection{Lower bound}
We will now derive a lower bound for the concurrence of $\hat \rho^{1,1}$ using the probability
\begin{align}
P_{\up,\up}\left(\theta\right)\defs P_{n^+_{A}=1,n^-_{A}=0,n^+_{B}=1,n^-_{B}=0}(\theta)
\end{align}
of finding a spin-up atom on each of the sites $A$ and $B$. The reason for focusing on $P_{\up,\up}\left(\theta\right)$ will become apparent in Sec.~\ref{analysis}. Using the rotation formula, we find the important result
\begin{align}
P_{\up,\up}(\pi/2)=\frac{p^{1,1}}{4}+\frac{1}{2}\Re[\rho_{\up\down}], \label{myreal}
\end{align}
where $\Re$ denotes the real part. The key point is that one can still detect $\Re[\rho_{\up\down}]$ in the presence of number fluctuations using
\begin{align}
\Re[\rho_{\up\down}]=2\left(P_{\up,\up}(\pi/2)-\frac{p^{1,1}}{4}\right)\nonumber.
\end{align}
With the same reasoning as in Sec.~\ref{spinresult}, we find the lower bound for the concurrence in the spin-$1/2$ sector
\begin{align}
G(\hat\rho_{A,B})&\leq C(\rho^{1,1}_{A,B}), \nonumber\\
G(\hat\rho_{A,B})&\defs 4 \left|P_{\up,\up}\left(\frac{\pi}{2}\right)-\frac{p^{1,1}}{4}\right| - 2\sqrt{P_{\up,\up}(0)P_{\up,\up}(\pi)}.\label{improved_bound}
\end{align}
\section{Scheme in the presence of number fluctuations without full spin-resolution}\label{nospin_resolution}
\subsection{Observable}
Current implementations of single-site resolved imaging in optical lattices do not resolve the individual atom numbers of both species~\cite{Fukuhara:2013}. Instead, the procedure is to push out one of the species using a resonant pulse and to detect the remaining atoms. For concreteness, we assume that the minus component is pushed out. The observed probability for the atom numbers $n^+_{A},n^+_{B}$ of the remaining plus-atoms is then
\begin{align}
\bar P_{n^+_{A},n^+_{B}}(\theta)\defs \sum_{n^-_{A},n^-_{B}}P_{n^+_{A},n^-_{A},n^+_{B},n^-_{B}}(\theta). \label{prob_push_out}
\end{align}
The detected signal therefore mixes contributions from different minus-atom numbers.\\
In addition to the probabilities after push-out, one can also simply image without push-out pulse. The observed probability thus corresponds to measuring the probability $p^{n_A,n_B}$  for the total atom numbers $n_{A},n_{B}$ according to Eq.~(\ref{pnanb}).
\subsection{Analysis of the problem}\label{analysis}
A key obstacle for formulating a lower bound without spin-resolution is to extract $\Re[\rho_{\up\down}]$ from the detected signal $\bar P_{n^+_{A},n^+_{B}}(\theta)$. For deriving bounds for the concurrence, we will use $\bar P_{1,1}(\theta)$, which does not contain a signal from empty lattice sites.
 Writing out Eq.~(\ref{prob_push_out}), we find
\begin{align}
\bar P_{1,1}(\pi/2)&=\frac{1}{2}\Re[\rho_{\up\down}]+\frac{p^{1,1}}{4}\nonumber\\
&+\sum_{n_A^->0,n_B^->0} P_{1,n_A^-,1,n_B^-}(\pi/2).\label{bad_contributions}
\end{align}
Using the rotation formula~(\ref{rot_general}), we can express the unwanted contributions in the second line in terms of probabilities before the pulse, $P_{n_A^+,n_A^-,n_B^+,n_B^-}(0)$, for states with at least one of the sites occupied by two or more atoms of the same species.\\
Consequently, these terms vanish for fermionic atoms in a single band Hubbard model~\cite{Esslinger:2010}. A suppression of doubly occuped sites for bosons is possible if the local chemical potential $\mu$ in optical lattice experiments is tuned to lower values ($0\lesssim \mu\lesssim 0.5 U$) at the expense of increasing the probability for holes~\cite{Sherson:2010}. Further, for experiments with Rydberg atoms in optical tweezers~\cite{Gaetan:2009,Urban:2009,Wilk:2010,Isenhower:2010,Gaetan:2010}, the filling of the traps is typically only zero or one.\\
In these situations, the terms in the second line of Eq.~(\ref{bad_contributions}) vanish and the bound (\ref{improved_bound}) can still be used without full spin resolution. For situations when doubly occupied sites of the same species cannot be neglected, modified bounds can be found by making certain assumptions on $\hat \rho_{A,B}$. We outline two methods in the following sections.
\subsection{Lower bound based on subtraction of $1/4$}
A modified version of the bound (\ref{improved_bound}) can be derived, using the following assumptions:
\begin{itemize}
\item {\bf{A1}} The probability of finding sites occupied by three or more atoms before applying the pulse is negligible: $P_{n^+_{A},n^-_{A},n^+_{B},n^-_{B}}(0)\approx 0$ if $n^+_{A}+n^-_{A}\geq 3$ or $n^+_{B}+n^-_{B}\geq 3$.
\item {\bf{A2}} There is no coherence between a state with two minus-atoms on $A$ and two plus-atoms on $B$ and a state with two plus-atoms on $A$ and two minus-atoms on $B$: $\rho_{\s{\bar n^+_A=2,\bar n_A^-=0, \bar n_B^+=0,\bar n_B^-=2\\n^+_A=0,n_A^-=2,n_B^+=2,n_B^-=0}}\approx 0$.
\end{itemize}
Assumption {\bf{A1}} is well fulfilled in the deep Mott-insulating regime of the Bose-Hubbard model at unity average filling for realistic experimental temperatures $T\approx 0.1 U$ \cite{Sherson:2010, Fukuhara:2013}. A breakdown of the second assumption {\bf{A2}} would require a non-negligible probability for having one of the sites occupied by two minus-atoms and the other site occupied by two plus-atoms. Again, for the Bose-Hubbard model at unity average filling for realistic experimental temperatures~\cite{Sherson:2010,Fukuhara:2013}, the joint probability of having both sites doubly occupied (independent of the spin) is much lower than one percent. Hence, assumption {\bf{A2}} is typically valid.\\
Using {\bf{A1}} and {\bf{A2}}, one can show that 
\begin{align}
2\left(\bar P_{1,1}(\pi/2)-\frac{1}{4}\right) &\leq \Re[\rho_{\up\down}]\nonumber.
\end{align}
Using the fact that $
\bar P_{1,1}(0)\geq P_{\up,\up}(0)$ and  $\bar P_{1,1}(\pi)\geq P_{\down,\down}(0)$, we arrive at a corresponding bound
\begin{align}
\bar  G_b(\hat \rho_{A,B})&\leq C(\hat\rho^{1,1}_{A,B}),\nonumber\\
\bar G_b(\hat \rho_{A,B})&\defs 4\left(\bar P_{1,1}(\pi/2)-\frac{1}{4}\right)-2\sqrt{\bar P_{1,1}(0)\bar P_{1,1}(\pi)}. \label{stupid_bound}
\end{align}
The bound (\ref{stupid_bound}) works with rather weak assumptions but is not particularly tight. The reason is that the subtraction of $\frac{1}{4}$ instead of $\frac{p^{1,1}}{4}$ leads to a reduction of the bound when $p^{1,1}$ is significantly smaller than one. Additionally, there is no absolute value around the first term, which can lead to a negative signal if $\Re[\rho_{\up\down}]<0$.
\subsection{Lower bound based on correlations}
 Therefore, we derive an improved bound compared to Eq.~(\ref{stupid_bound}) based on evaluating the quantity
\begin{align}
\bar P^c_{1,1}(\theta)\defs\bar P_{1,1}(\theta)-\bar P_{1,A}(\theta)\bar P_{1,B}(\theta)\nonumber,
\end{align}
where $\bar P_{1,j}(\theta)$ are the single-site probabilities for observing a single up-spin atom on site $j=A$  or $B$ after push-out of the minus-component. They are related to the joint probability $\bar P_{n^+_{A},n^+_{B}}(\theta)$ via
\begin{align}
\bar P_{1,A}(\theta)=\sum_{n_B} \bar P_{1,n_B}(\theta)\nonumber\\
\bar P_{1,B}(\theta)=\sum_{n_A} \bar P_{n_A,1}(\theta). \label{single_site prob}
\end{align}
The subscript $c$ for $P^c_{1,1}(\theta)$ stands for 'connected' because $P^c_{1,1}(\theta)$ resembles the form of a connected correlation function.\\
Using Eq.~(\ref{myreal}), we find
\begin{align}
&\bar P^c_{1,1}(\pi/2)=\frac{1}{2}\Re[\rho_{\up\down}]+\frac{p^{1,1}-p^{1,A}p^{1,B}}{4}\nonumber\\
&+\sum_{n_A^->0,n_B^->0}(P_{1,n_A^-,1,n_B^-}(\pi/2)-P_{1,n_A^-}(\pi/2)P_{1,n_B^-}(\pi/2)) \label{connected1},
\end{align}
where, for $j=A$ or $B$, $p^{1,j}$ is the single-site probability of finding the total atom number $n_{j}=1$ and $P_{1,n_{j}^-}(\theta)$ is the probability of finding a single plus-atom and $n_{j}^-$ minus-atoms after a $\theta$-pulse. The connection to the corresponding joint probabilities is analogous to Eq.~(\ref{single_site prob}).\\
The signal $P^c_{1,1}(\pi/2)$ yields $\Re[\rho_{\up\down}]$ via
\begin{align}
2 \bar P^c_{1,1}(\pi/2)\approx \Re[\rho_{\up\down}] \label{connected_signal}
\end{align}	
if the following two assumptions hold:
\begin{itemize}
\item {\bf{B1}} The probability of finding a single atom on site $A$ is independent of the probability of finding a single atom on site $B$: \mbox{$p^{1,1}\approx p^{1,A}p^{1,B}$}.
\item {\bf{B2}} There are no correlations in the sectors with local occupation number higher than one: \mbox{$\hat \rho^{n_A,n_B}_{A,B}\approx \hat \rho^{n_A}_A \otimes \hat \rho^{n_B}_B$ for $n_A,n_B>1$.}
\end{itemize}
With $\rho^{n_j}_j$ we refer to the single-site reduced density operator for site $j$ projected onto local atom number $n_j$ (see Eq.~(\ref{number_projected})).\\
A mechanism that would violate these assumptions is the introduction of density-density correlations via quantum fluctuations in the form of particle-hole pairs~\cite{Endres:2011}. However, these correlations are extremely small beyond nearest-neighbor distances. A potential danger arises if the system is brought out of equilibrium, for example, via a fast quench, which can induce longer-range density-density correlations~\cite{Cheneau:2012}. This can be avoided with a careful adjustment of lattice ramps.\\
Additionally, there are experimental checks for the validity of Eq.~(\ref{connected_signal}), such as an observation of the checkerboard pattern described in Sec.~\ref{quality} on a zero background signal, that is, $\bar P^c_{1,1}(\pi/2)\approx 0$ for even distances. Further, the absence of density-density correlations can be checked using imaging without a push-out pulse.\\
Based on Eq.~(\ref{connected_signal}) we can formulate a lower bound
\begin{align}
\bar G_c(\hat\rho_{A,B}) &\leq C(\rho^{1,1}_{A,B})\nonumber,\\
\bar G_c(\hat\rho_{A,B})&\defs 4 \left|\bar P^c_{1,1}(\pi/2)\right| - 2\sqrt{\bar P_{1,1}(0)\bar P_{1,1}(\pi)}, \label{improved_bound2}
\end{align}
which holds if the assumptions {\bf{B1}} and {\bf{B2}} are fulfilled.
\subsection{Influence of parity-projection}\label{parity}
In the current experiments with single-site resolution, only the parity of the on-site occupation number can be observed due to a pair-wise loss from light-assisted collisions~\cite{Bakr:2010,Sherson:2010}. The parity-projection only occurs during the actual detection of the remaining species but not during the push-out~\cite{Fukuhara:2013}. The observed probabilities after push-out and subsequent parity projection are thus
\begin{align}
\tilde P_{n^+_{A},n^+_{B}}(\theta)\defs \!\!\!\! \!\!\!\! \!\!\!\! \!\!\!\! \!\!\!\! \!\!\!\!\sum_{\s{\bar n^+_{A},\bar n^-_{A},\bar n^+_{B},\bar n^-_{B} \\ \bar n^+_A \!\!\!\!\!\!\mod_{2}=n^+_{A},\,\bar n^+_B\!\!\!\!\!\! \mod_{2}=n^+_{B}} } \!\!\!\!\!\!\!\!\!\!\!\!\!\!\!\!\!\!\!\!P_{\bar n^+_{A},\bar n^-_{A},\bar n^+_{B},\bar n^-_{B}}(\theta), \nonumber
\end{align}
where $n^+_{A},n^+_{B}<2$.\\
The additional terms that enter $\tilde P_{1,1}(\theta)$ all stem from triply or higher occupied sites. These terms vanish if {\bf{A1}} is fulfilled. Consequently, the bound (\ref{stupid_bound}) can still be used with parity-projection.\\
Similarly, parity-projection adds several terms to Eq.~(\ref{connected1}) which all vanish if {\bf{B2}} holds. Therefore, bound (\ref{improved_bound2}) also remains unaffected.\\

\section{Conclusion and outlook}
In conclusion, we proposed a scheme for detecting lower bounds for the concurrence of two sites of a lattice many-body system, which could be used for measuring spin-entanglement in quantum magnetism experiments with coexisting spin and number fluctuations.\\
Our analysis showed that a detection of the lower bounds should be possible in current high-resolution imaging setups for quantum gases in optical lattices~\cite{Bakr:2009, Bakr:2010, Sherson:2010, Endres:2011,Simon:2011,Fukuhara:2013, Fukuhara:2013b} despite several technical limitations. However, the scheme would simplify if full spin-resolution was achieved experimentally, and the bound (\ref{improved_bound}) could be used. \\
A possible solution for one-dimensional systems is to prepare a single chain of atoms and let the atoms tunnel orthogonally to the chain before the detection. If a magnetic field gradient is applied during the orthogonal dynamics, atoms with positive and negative magnetic moment would spatially separate. The spatial separation could allow a detection of the local occupation numbers of both spin states in a single experimental run. In this sense, an in-situ Stern-Gerlach experiment could be realized with full spatial resolution along the one-dimensional chain. Such a scheme could also be useful to detect the correlations induced by impurities in strongly interacting superfluids, enabling the direct imaging of a polaron cloud~\cite{Fukuhara:2013}.\\
Concerning the actual influence of on-site number fluctuations on spin-entanglement, we performed numerical simulations of spin impurity dynamics in the one-dimensional Bose-Hubbard model. The effect of quantum fluctuations within large parts of the Mott insulating phase could be captured by a renormalized XX-spin dynamics normally only valid in the very deep Mott insulating regime. A similar behavior results from thermally activated number fluctuations. Importantly, our simulations showed that the entanglement generation and spreading survives for the temperatures and parameters of current experiments~\cite{Fukuhara:2013, Fukuhara:2013b}. Thus, the application of the proposed detection technique for this type of experiment should be immediately possible. More generally, the experimental detection of spin-entanglement in Hubbard models realized with optical lattices~\cite{Kuklov:2003, Duan:2003, Trotzky:2008a, Simon:2011,Fukuhara:2013, Fukuhara:2013b,Greif:2013,Nascimbene:2013} is now within reach.\\
\section*{Acknowledgments}
We acknowledge indispensable discussions with T. Fukuhara, C. Gross, I. Bloch, and G. Giedke.
This work was supported by EU (IP-SIQS), 
by Italian MIUR via PRIN Project 2010LLKJBX and via FIRB Project RBFR12NLNA, 
and by Regione Toscana POR FSE 2007-2013. 
\bibliographystyle{h-physrev}
\bibliography{bibliography}
\end{document}